\documentclass[11pt]{article}
\usepackage{graphicx}
\usepackage{amssymb}
\def\lsim{\mathrel{\raise.3ex\hbox{$<$\kern-.75em\lower1ex\hbox{$\sim$}}}}
\def\gsim{\mathrel{\raise.3ex\hbox{$>$\kern-.75em\lower1ex\hbox{$\sim$}}}}

\begin{document}

\title{Classification of dark energy models in the ($w_0, w_a$) plane}

\author{V. Barger$^{1}$, E. Guarnaccia$^{2}$ and D. Marfatia$^{2}$\\[2ex]
\small\it $^1$Department of Physics, University of Wisconsin, Madison, WI 53706\\
\small\it $^2$Department of Physics and Astronomy, University of Kansas, Lawrence, KS 66045}

\date{}

\maketitle

\begin{abstract}
We classify dark energy models in 
a plane of observables that correspond to
the common parameterization of a non-constant equation of state, 
$w(a)=w_0+w_a(1-a)$, where $a$ is the scale factor of the universe.
The models fall into four classes and only two of these
classes have a region of overlap in the observable plane. 
We perform a joint analysis of all Type Ia supernova (SNIa) data 
compiled by the 
High-Z SN Search Team (HZT) and the 
Supernova Legacy Survey (SNLS) and find that no class of
models is excluded by current SNIa data. However, an analysis of 
large scale structure, Ly$\alpha$ forest and bias constraints from 
SDSS, the Gold SNIa data and WMAP data indicates that non-phantom barotropic
models with a positive sound speed 
are excluded at the 95\%~C.~L.

\end{abstract}

\newpage


Abundant cosmological 
data indicate that the expansion of the universe has changed
from a decelerating phase to an accelerating phase in the last few
billion years. This is generally attributed to the recent dominance
of an energy component with negative pressure called dark 
energy{\footnote{Modifications to the
Friedman-Robertson-Walker equation can also provide an explanation for the
current 
accelerated expansion such as in Cardassian models~\cite{katie} which
do not have exotic forms of energy or a vacuum contribution. We do not
consider such models here.}}.
For reviews see Ref.~\cite{Peebles:2002gy}.

Among candidates for dark energy are the cosmological constant $\Lambda$ 
proposed
by Einstein and a dynamical scalar field such as 
quintessence~\cite{Peebles:1987ek,frieman,coble}. 
The cosmological constant arises in particle physics as vacuum energy
with constant energy density $\rho$, constant pressure $p$ and equation
of state $w\equiv p/\rho=-1$. A cosmological constant is troublesome from
the particle physics standpoint because the cosmologically measured value
of $\rho_\Lambda^{1/4}$ is found to be about 
$(0.7\rho_c)^{1/4} \simeq 0.002$ eV, where $\rho_c$ is the critical density
for which the universe is flat.
Quantum field theory, however, suggests a value at least
15 orders of magnitude larger under the assumption that the
Standard Model is an effective theory valid below 1 TeV. For the vacuum 
energy to be renormalized to $0.002$ eV, a fine tuning in the energy density 
at the level 60 decimal
places is required. See 
Ref.~\cite{Weinberg:1988cp} for a review of the cosmological constant 
problem. Compounding this, is the cosmic coincidence problem which
seeks an explanation for why the
dark matter and vacuum energy densities are comparable today although
their ratio scales as $1/a^3$, where $a=1/(1+z)$ in terms of redshift 
$z$~\cite{Peebles:2002gy}. Quintessence is
a not a solution, but provides 
a phenomenological explanation for these problems.
It is a time-varying field with $w(a)>-1$ and is usually represented 
by a very light 
scalar field rolling down a potential. The potential and field values 
are chosen so that the energy density in quintessence dominates the dark 
matter density only recently and takes on the measured value. This picture
has been generalized to schemes with $w<-1$ called phantom dark energy and  
to dark energy as a barotropic fluid with $w=f(\rho)/\rho$. In what follows,
we consider  
phantom models, barotropic
fluid models of dark energy and two classes of quintessence models.

Recently, in a sequence of papers~\cite{Caldwell:2005tm,Scherrer:2005je,
Chiba:2005tj}, models of dark energy have been divided into 
categories depending on their equation of state, $w$, and 
its derivative with respect to the logarithm of the scale factor, 
$w^{\prime}\equiv {\frac{dw}{d \ln(a)}}$; since $H^{-1}= 
{\frac{dt}{d \ln(a)}}$, $w^{\prime}$ is the time derivative of
$w$ in units of Hubble time. It was pointed out that
different classes of models evolve in different regions of the
 $w-w^\prime$ plane{\footnote{Equivalently, different dark energy models 
evolve in different regions of statefinder planes~\cite{statefinder};
the statefinder diagnostic, defined in terms of the second and third time
derivatives of $a$, is directly related to $w$ and 
$w^\prime$.}}.

Building on this work, we classify models of dark energy in a
parameter space of observables. 
It has been argued in Ref.~\cite{Linder:2005ne}, 
that given the sensitivities of future
cosmological experiments, one can realistically 
expect to constrain at most two
parameters related to the equation of state. Specifically, we use the standard
parameterization of dark energy evolution in terms
of the equation of state and its derivative with respect to 
$z$ at the present time~\cite{Linder:2002et}:  
\begin{equation}
w(z) = w_0 + w_a {\frac{z}{1+z}}\,
\label{eos}
\end{equation}
with
\begin{equation}
w_a = {\frac{dw}{dz}}\big\arrowvert_{z=0} = -2 w^{\prime}\arrowvert_{z=1}\,.
\end{equation}
Note the relationship between $w_a$ and the value of $w^{\prime}$ at
$z=1$, the epoch when the dark energy contribution to the expansion rate 
is expected to increase in importance{\footnote{In 
Refs.~\cite{Riess:2004nr,Clocchiatti:2005vy}, the transition 
from matter domination to dark energy domination is found to occur at 
$z \simeq 0.5$. For a recent discussion of the low redshift evolution of the 
dark energy density 
see Ref.~\cite{low}.}. We classify models in the $w_0-w_a$ plane. 
The primary advantage of 
this parameterization is that since the 
dark energy density evolves according to
\begin{equation}
\rho(z)=\rho(0) (1+z)^{3(1+w_0+w_a)} e^{-3w_a {\frac{z}{1+z}}}\,,
\end{equation}
it is
well-behaved from high redshifts until today. Note that for 
dark energy to be subdominant
at early times, $w_0+w_a<0$. We do not impose the latter restriction since
the parameterization of Eq.~(\ref{eos}), though well-behaved at 
high redshift,
may be inapplicable there.

The previously popular 
linear parameterization $w(z)= w_0+w_1z$~\cite{oldparam} diverges at high 
redshift and consequently yields artificially strong constraints
on $w_1$ in analyses involving data at high redshifts such as CMB data.
Since Eq.~(\ref{eos}) 
is often used in the analysis of data,
our classification will enable a direct comparison between the results
of such analyses and the theoretical space of models. 

The categorization we develop is analogous to that of inflationary
models where models populate regions in a plane defined by 
the spectral index of scalar perturbations, $n_s$, and the relative 
normalization of the tensor and scalar spectra, 
$R$~\cite{Dodelson:1997hr}, or in a plane of
horizon-flow parameters~\cite{Barger:2003ym}.
\vskip 0.3in
{\bf{Dark energy models:}}

\begin{figure}[t]
\centering\leavevmode
\includegraphics[width=5.5in]{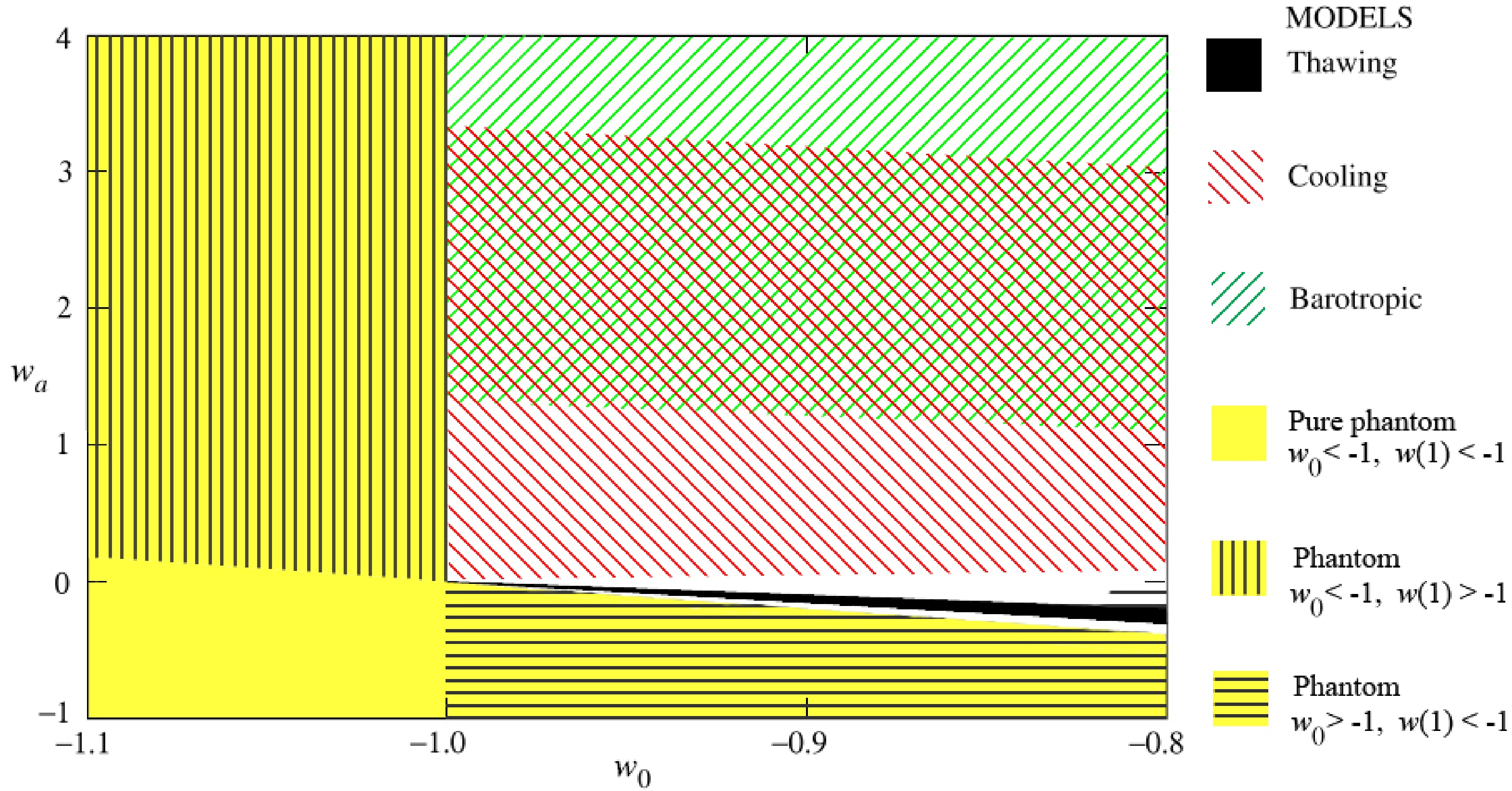}
\caption[]{Classification of the four types of models in the ($w_0,w_a$) 
plane. All regions depicted by a light yellow shaded region correspond
to phantom models with $w<-1$ today or recently.
\label{fig:fig1}}
\end{figure} 

Here, we briefly review the models under consideration, define the
regions they occupy in the $w-w^\prime$ plane and classify them in the 
$w_0-w_a$ plane. 

We first consider models which obey the null energy condition,
$w \ge -1$, and then consider phantom models with
$w < -1$.
\vskip 0.2in
{\it{\underline{Thawing models}}}

The nomenclature, ``thawing'' models, was coined in 
Ref.~\cite{Caldwell:2005tm} to describe
a scalar field whose equation of state increases (thaws out) 
from $w \simeq -1$ as the scalar rolls down towards the
minimum of its potential. Potentials of the form $\phi^n$ (with $n>0$) 
and $e^{-\phi}$ are typical of these models. 
Particle-physics models involving axions~\cite{axion}, 
pseudo Nambu-Goldstone 
bosons~\cite{frieman}, moduli or dilatons~\cite{Barreiro:1999zs} 
often have such potentials.
They are found to satisfy~\cite{Caldwell:2005tm} 
\begin{equation}
1+w < w^{\prime} < 3 (1+w)\,.
\label{thawing}
\end{equation}
It is natural to impose this constraint at the 
epoch when dark energy starts becoming
important, {\it i.e.}, $z=1$.
With $w(1)=w_0+w_a/2$ and  
$w^{\prime}\arrowvert_{z=1}=-w_a/2$, we find
\begin{equation}
-{\frac{3} {2}}(1+w_0) < w_a < - (1+w_0)\,.
\label{thawing1}
\end{equation}
Thawing models occupy the dark-shaded triangular wedge in 
Fig.~\ref{fig:fig1}.

\vskip 0.2in
{\it{\underline{Cooling models}}}

In these models, initially, $w > -1$ and $w$ decreases as 
the scalar rolls down the potential, which is typically of the
form $\phi^{-n}$ or $\phi^{-n} e^{\phi^2}$ (with $n>0$). 
The tracker models of 
Ref.~\cite{Zlatev:1998tr} adopt such potentials. Such forms of the potential 
also arise in
models of dynamical supersymmtery 
breaking~\cite{Binetruy:1998rz} and
supergravity~\cite{Brax:1999gp}.
These models lie in a region of
the $w-w^\prime$ plane defined 
by~\cite{Caldwell:2005tm,Scherrer:2005je,Chiba:2005tj}
\begin{equation}
-3(1-w)(1+w)< w^{\prime} < 0.2 w (1+w)\,.
\label{cooling}
\end{equation}
By requiring that $w(1)$ obey the above inequality, we obtain the region
with red slant hatches (with negative slope) in Fig.~\ref{fig:fig1}. 
As shown in Ref.~\cite{Chiba:2005tj}, 
$k$-essence models~\cite{Armendariz-Picon:2000dh}
with a non-linear kinetic term also 
fall within the class of cooling models.

Note that the ``freezing'' models of Ref.~\cite{Caldwell:2005tm} 
are a special case 
corresponding to the situation in which the potential has a minimum
at $\phi=\infty$. For freezing models,
\begin{equation}
3w(1+w)< w^{\prime} < 0.2 w (1+w)\,.
\label{freezing}
\end{equation}

\vskip 0.2in
{\it{\underline{Barotropic fluids}}}

A barotropic fluid is one for which the density depends only on the 
pressure. Supposing that dark energy is a barotropic fluid with 
$p=f(\rho)$ leads to a class of models that contains as special
cases the generalized Chaplygin gas~\cite{Bento:2002ps} for which 
$f(\rho)= -A/\rho^\alpha$ with $\alpha >-1$
and the orginal Chaplygin gas model with 
$\alpha=1$~\cite{Kamenshchik:2001cp}. While the models of
Refs.~\cite{Bento:2002ps,Kamenshchik:2001cp} were proposed to unify
dark matter and dark energy, subsequent work addressed the possibility
of Chaplygin gas models as being models of dark energy 
alone~\cite{Dev:2002qa}. Barotropic fluid 
models arise in string theories~\cite{strings} 
in which
{\it {e.g.}} the Chaplygin gas corresponds to a gas of $d$-branes in a
$d+2$ spacetime~\cite{strings2}.

Non-phantom barotropic models were found to satisfy~\cite{Scherrer:2005je}
\begin{equation}
w^{\prime} < 3 w (1+w)\,
\label{barotropic}
\end{equation}
under the assumption that $c_s^2 \equiv dp/d\rho >0$ so that perturbation
growth is stable.

The region with green
slant hatches (with positive slope) 
in Fig.~\ref{fig:fig1} depicts barotropic fluids.
This region partially overlaps with that 
for cooling models.

Comparing Eq.~(\ref{freezing}) with Eq.~(\ref{barotropic}), we see that
freezing models and barotropic fluids have nonoverlapping regions
with a common boundary in the 
$w-w^{\prime}$ plane. In the observable $w_0-w_a$ plane, freezing models
occupy the region with red slant hatches (with negative slope) 
below the cross hatched region (non-overlapping
 with the region occupied by barotropic fluid models).
See Fig.~\ref{fig:fig1}.
\vskip 0.2in
{\it{\underline{Phantom models}}}

Finally, we consider phantom models for which $w<-1$ today or recently.  
It was shown in Ref.~\cite{Chiba:2005tj} 
that for the prototypical phantom ({\it i.e.} 
a ghost with a negative kinetic term) model~\cite{Caldwell:1999ew},
\begin{equation}
3(1-w)(1+w) < w^{\prime} < 3w(1-w)(1+w)\,.
\label{phantom}
\end{equation}
However, we do not impose this inequality since it 
is very marginally more constraining than
$w< -1$, and since the upper bound on $w^{\prime}$ was obtained 
specifically for tracker phantom models.
Requiring $w_0<-1$ or $w(1)<-1$ we find that phantom 
models populate the light yellow shaded region in Fig.~\ref{fig:fig1}.

The subregion defined by vertical (horizontal) lines corresponds to 
$w_0<-1$ and $w(1)>-1$ ($w_0>-1$ and $w(1)<-1$) 
{\it i.e.,} models for which the equation of state crossed the phantom
divide line $w=-1$ from a higher value to a lower value (lower value to
a higher value). The light yellow shaded region without lines corresponds
to ``pure phantom'' models for which the phantom divide line has not been
crossed recently {\it i.e.,} $w_0<-1$ and $w(1)<-1$.

Note that not all models that violate the null
energy condition require the scalar to be a phantom. Demonstrations
of the violation of the null energy condition have been made in
models of vacuum metamorphosis~\cite{Parker:2001ws}, 
climbing scalar fields~\cite{Csaki:2005vq}, 
and the braneworld~\cite{Lue:2004za} without 
the introduction
of negative energies or negative norm states.

Interestingly, the quintom model of Ref.~\cite{quintom} lies in the region with
vertical lines.

\vskip 0.2in

We caution against interpreting this classification as being comprehensive. 
For example, dark energy with an oscillating equation of 
state~\cite{Barenboim:2004kz}
can rarely be realized in a potential formulation.
(An example of oscillating dark energy from a quite complicated potential 
can be found in Ref.~\cite{Barenboim:2005np}).
Thus, oscillating dark energy as a class of models 
falls outside the realm of 
our classification unless the equation of state crosses the phantom divide
line as in quintom models~\cite{quintom}. 
Similarly, we have not attempted to include neutrino 
dark energy~\cite{dark-energy} or dark energy with generalized 
equations of state~\cite{generalized} in the classification.

When viewed in conjunction with data, 
our classification is intended to provide 
guidance as to what kinds of schemes are preferred by data. 
It is necessary that a cosmological constant $(w_0,w_a)=(-1,0)$,
 be excluded by data to
distinguish between thawing, cooling and phantom models because these
models have a cosmological constant as a limiting case.
This is in contrast to barotropic fluids with a positive sound speed.

\vskip 0.3in
{\bf{Current status:}}

Having defined the regions that the four different types of models 
occupy in the $w_0-w_a$ plane, we now examine if current data are able
to discriminate between these classes. 

We perform a joint analysis of SNIa data compiled by
 the High-Z SN Search Team (HZT) and the Supernova Legacy Survey (SNLS). 
We include
the Gold and Silver datasets of Ref.~\cite{Riess:2004nr}, 
the 4 SN of Ref.~\cite{Clocchiatti:2005vy}, and the SNLS
dataset~\cite{Astier:2005qq} assuming a flat universe with 
$0.2 \leq \Omega_m \leq 0.4$ and $0.5 \leq h \leq 0.9$.
We have removed the SN common to both surveys from the SNLS
dataset which results in a full dataset of 266 SN.  
 Our joint analysis
is implemented as follows.
The statistical significance of a cosmology is determined in terms of
$\chi^2= \chi^2_{\rm{HZT}} +\chi^2_{\rm{SNLS}}$, with
\begin{eqnarray}
\chi_{\rm{HZT}}^2 &=& \sum_{i=1}^{188} {\frac{(\mu_i^{obs}-5\log_{10} 
d_L(z_i; \Omega_m, h, w_0, w_a)-M_1)^2}{\sigma_i^2}} \,, \\
\chi_{\rm{SNLS}}^2 &=& \sum_{i=1}^{78} {\frac{(\mu_i^{obs}-5\log_{10} 
d_L(z_i; \Omega_m, h, w_0, w_a)-M_2)^2}{\sigma_{\mu_i}^2+\sigma_{int}^2}} \,.
\end{eqnarray}
where $\mu_i^{obs}$ is the distance modulus at redshift $z_i$, 
$d_L$ is the luminosity distance (in units of 10 pc), $\sigma_i$ is the total
uncertainty in the distance modulus,
$\sigma_{\mu_i}^2$ is
the measurement variance and
$\sigma_{int}$ is the intrinsic dispersion of SN absolute magnitudes. 
$M_1$ and $M_2$ are nuisance parameters that are marginalized
over in the fit. They correspond to analysis-dependent 
global unknown constants 
in the definition of distances. $M_1$ and $M_2$ are analogous to absolute 
magnitudes if observational data are provided
as apparent magnitudes. 

Note that SNLS provides the apparent magnitudes,
the stretch factor used to calibrate them, and the rest frame color factor
that measures host galaxy dust extinction. We have chosen to use 
the derived $\mu$ values instead for two reasons: (1) it is not possible 
to perform an analysis at the same level of sophistication for the HZT
data since the High-Z SN Search Team provides only the distance modulus
({\it{i.e.,}} after corrections have been made to the apparent magnitude 
$m(z)$), and (2) it would be overkill considering 
that current SN data are not very constraining in the $(w_0,w_a)$ plane.

\begin{figure}[t]
\centering\leavevmode
\includegraphics[width=4.5in]{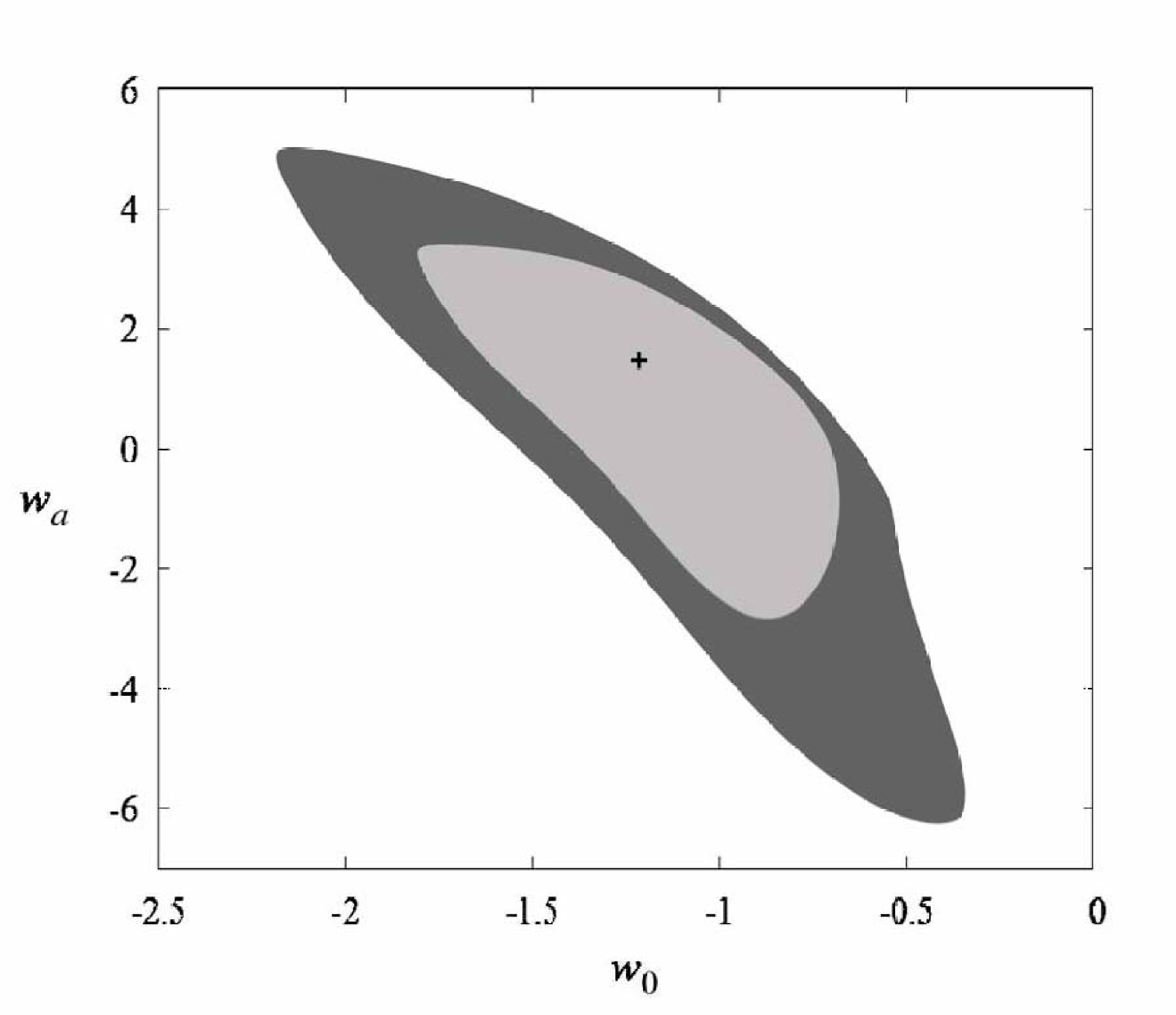}
\caption[]{68\% C.~L. and 95\% C.~L. allowed 
regions from a combined analysis of
the HZT and SNLS SNIa data. The crosshair marks the best-fit point 
$(w_0,w_a)=(-1.2,1.5)$. No class of dark energy model is excluded.
\label{fig:fig2}}
\end{figure}

Reference~\cite{Astier:2005qq} 
finds $\sigma_{int}=0.13\pm 0.02$. 
To be conservative in our combined analysis we choose $\sigma_{int}=0.15$.
We emphasize that our approximate analysis of SNLS data alone yields good 
agreement with that of the more accurate analysis of 
Ref.~\cite{Astier:2005qq}
implemented in Ref.~\cite{Nesseris:2005ur}. Gravitational lensing bias
could be reduced by use of the flux-averaging procedure of 
Ref.~\cite{wang}.

In Fig.~\ref{fig:fig2} we display the
68\% and 95\% C.~L. allowed regions obtained in our analysis. 
We immediately conclude
that all classes of dark energy models are comfortably allowed by current
SN data. 
In Fig.~\ref{fig:fig3} we overlay the results of an analysis of 
large scale structure, Ly$\alpha$ forest and bias constraints from 
SDSS, the Gold SNIa data and WMAP data on the plane of classification. 
The 68\% and 95\% C.~L. allowed 
regions have been adapted from Fig.~9 of Ref.~\cite{Seljak:2004xh} 
after accounting for 
the fact that our $w_a$ corresponds to $-w_1$ in Ref.~\cite{Seljak:2004xh}.

It appears that current data exclude non-phantom 
barotropic models with a positive sound speed at the 
95\% C.~L. Also, among cooling models, freezing models appear to be favored.
However, it may be too soon to draw these conclusions with 
confidence because the analysis of several correlated datasets has 
led to these results. It will be more convincing if the same conclusions
can be made from an analysis of fewer distinct but larger datasets. 
For a recent assessment of the abilities of future dark energy surveys  
to discover the 
 time evolution of $w$ see Ref.~\cite{liddle}.

\begin{figure}[th]
\centering\leavevmode
\includegraphics[width=5.5in]{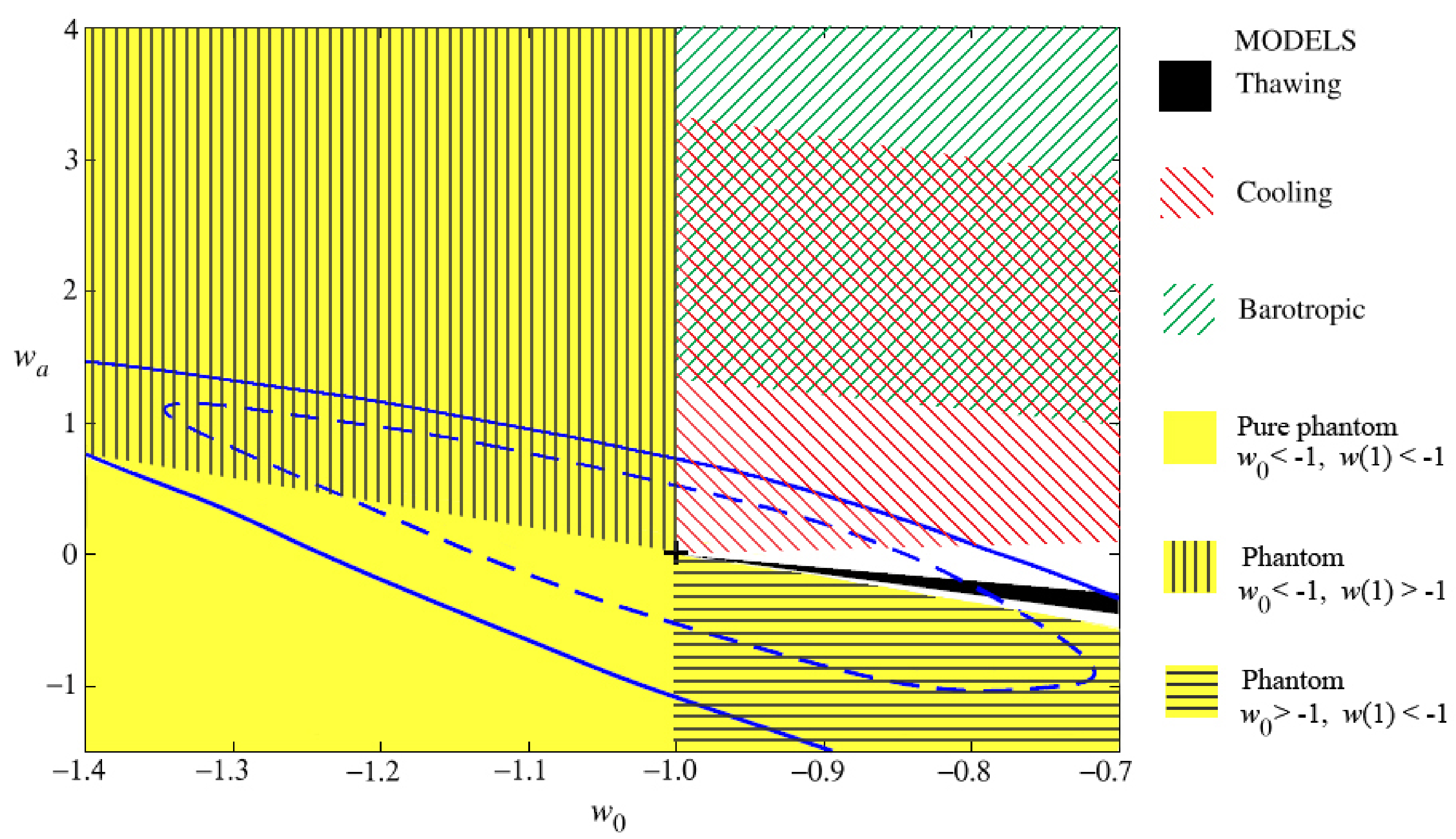}
\caption[]{68\% C.~L. and 95\% C.~L. allowed regions from an analysis of
large scale structure, Ly$\alpha$ forest and bias constraints from
SDSS, the Gold SNIa data and WMAP data. The crosshair marks the 
best-fit point $(w_0,w_a)=(-1,0)$, which corresponds to a 
cosmological constant.
 Non-phantom barotropic models
are excluded at the 95\% C.L. (Adapted from Fig.~9 of 
Ref.~\cite{Seljak:2004xh}).
\label{fig:fig3}}
\end{figure}

\newpage
\vskip 0.3in

{\bf{Acknowledgments:}}

We thank P.~Astier and Y.~Wang for helpful communications.
This research was supported by the U.S. Department of Energy
under Grant No.~DE-FG02-95ER40896, by the NSF
under CAREER Award No.~PHY-0544278 and Grant No.~EPS-0236913, 
by the State of Kansas through the
Kansas Technology Enterprise Corporation and by the Wisconsin Alumni
Research Foundation.



\end{document}